\documentclass[prb,aps,twocolumn,superscriptaddress,showpacs]{revtex4-1}
\usepackage[colorlinks=true,linkcolor=black, citecolor=blue, urlcolor=blue, 
unicode=true,breaklinks]{hyperref}

\usepackage{graphicx,mathtools}

\begin{document}

\title{\texorpdfstring{Acoustic lattice instabilities at the magneto-structural transition in Fe$_{1.057(7)}$Te}{}}

\author{K. Guratinder}
\affiliation{School of Physics and Astronomy, University of Edinburgh, Edinburgh EH9 3JZ, United Kingdom}

\author{E. Chan}
\affiliation{Institut Laue-Langevin, 71 avenue des Martyrs, CS 20156, 38042 Grenoble Cedex 9, France}
\affiliation{School of Physics and Astronomy, University of Edinburgh, Edinburgh EH9 3JZ, United Kingdom}

\author{E. E. Rodriguez}
\affiliation{Department of Chemistry and  Biochemistry, University of Maryland, College Park, Maryland 20742, USA}

\author{J. A. Rodriguez-Rivera}
\affiliation{NIST Center for Neutron Research, National Institute of Standards  and Technology, Gaithersburg, Maryland 20899-6100, USA}
\affiliation{Department of Materials Science, University of Maryland, College Park, MD  20742}

\author{U. Stuhr}
\affiliation{Laboratory for Neutron Scattering, Paul Scherrer Institut, CH-5232 Villigen, Switzerland}

\author{A. Stunault}
\affiliation{Institut Laue-Langevin, 71 avenue des Martyrs, CS 20156, 38042 Grenoble Cedex 9, France}

\author{R. Travers}
\affiliation{School of Physics and Astronomy, University of Edinburgh, Edinburgh EH9 3JZ, United Kingdom}

\author{M. A. Green}
\affiliation{School of Physical Sciences, University of Kent, Canterbury CT2 7NH, United Kingdom}

\author{N. Qureshi}
\affiliation{Institut Laue-Langevin, 71 avenue des Martyrs, CS 20156, 38042 Grenoble Cedex 9, France}

\author{C. Stock}
\affiliation{School of Physics and Astronomy, University of Edinburgh, Edinburgh EH9 3JZ, United Kingdom}

\date{\today}

\begin{abstract}
	
	Fe$_{1.057(7)}$Te undergoes a first-order structural transition from a high temperature tetragonal phase to a low temperature monoclinic at T$_{S} \sim 70$ K, breaking the fourfold C$_{4}$ high temperature lattice symmetry.  At the same temperature, time reversal symmetry is broken with magnetic iron ions ordering in a commensurate (with the nuclear lattice) bicollinear arrangement.  The low-energy magnetic dynamics proximate to this magneto-structural transition are, however, incommensurate and have been reported on previously [Phys. Rev. B {\bf{95}}, 144407 (2017)].  In this current work, we investigate the soft acoustic lattice dynamics near this combined magneto-structural transition.  We apply spherically neutron polarimetry to study the static magnetism near this transition, characterized with x-ray powder diffraction, and find no evidence of static incommensurate magnetic correlations near the onset of monoclinic and bicollinear antiferromagnetic order.  This fixes the position of our single crystal sample in the Fe$_{1+x}$Te phase diagram in the magnetic bicollinear region and illustrates that our sample statically undergoes a transition from a paramagnetic phase to a low-temperature bicollinear phase.  We then apply neutron spectroscopy to study the acoustic phonons, related to elastic deformations of the lattice.  We find a temperature dependent soft acoustic branch for phonons propagating along [010] and polarized along [100].  The slope of this acoustic phonon branch is sensitive to the elastic constant $C_{66}$ and the shear modulus.  The temperature dependence of this branch displays a softening with a minimum near the magneto-structural transition of T$_{S}$ $\sim$ 70 K and a recovery within the magnetically ordered low temperature phase.  Soft acoustic instabilities are present in the collinear phases of the chalcogenides Fe$_{1+x}$Te where nematic order found in Fe$_{1+\delta}$Se is absent.  We speculate, based on localized single-ion magnetism, that the relative energy scale of magnetic spin-orbital coupling on the Fe$^{2+}$ transition metal ion is important for the presence of a nematicity in the chalcogenides.

\end{abstract}


\maketitle

\section{Introduction}

The unexpected discovery of high-temperature superconductivity in a family of materials based on iron in 2008 \cite{Kamihara08:130,Hsu08:105} resulted in an intense research effort in the structural and magnetic properties of iron based chalcogenide and pnictide materials.~\cite{Johnston10:59,Lumsden10:22,Paglione10:6,Dai12:8,Dai15:87,Chubukov15:68,Fernandes22:601,Inosov16:17,Ishida09:78,Stock16:28,Mazin10:464}  The iron based compounds display strong magneto-structural coupling and materials proximate to superconductivity~\cite{Nandi10:104,Bohmer11:6} undergo both a structural transition (from high temperature tetragonal to low temperature orthorhombic phases) and the formation of a low temperature spin density wave phase.  Many such compounds display a high temperature region where the four-fold C$_{4}$ symmetry is broken through either a structural transition or an electronic~\cite{Kuo16:352} instability, however, no magnetic order is present preserving time reversal symmetry.~\cite{Fernandes14:10,Fernandes12:25}  The breaking of the fourfold symmetry is further illustrated through transport measurements of the resistivity.~\cite{Ishida20:117}  This unusual intermediate phase which lacks an ordered magnetic moment has been termed a ``nematic" phase~\cite{Chandra91:66,Fradkin10:1} in analogy to a similar symmetry broken phase present in liquid crystals and its close proximity to superconductivity~\cite{Fernandes14:111} has made it the focus of many investigations.  Important for the formation of the ``nematic" phase is the softening of a shear modulus which can be measured through the slope of acoustic phonons with scattering techniques.  In this report, we investigate the soft acoustic lattice dynamics in Fe$_{1.057(7)}$Te, a compound parent to chalcogenide unconventional superconductivity.

Arguably, the simplest of all iron pnictide and chalcogenide based compounds is the single-layer Fe$_{1+x}$Te~\cite{Chiba55:10,Fruchart75:10} which has \textit{not} been reported to display superconductivity. However, it is parent to superconductivity as anion substitution on the tellurium site with, for example, sulfur or selenium results in superconductivity.~\cite{Sales09:79,Mizuguchi09:94,Wen11:74,Zajdel10:132,Bendele10:82}  The chalcogenide Fe$_{1+x}$Te does not display an observable nematic phase, like its pnictide counterparts, with both the high temperature C$_{4}$ structural and magnetic time reversal symmetry being broken simultaneously at a common magneto-structural transition temperature.~\cite{Li09:79}  However, the closely related Fe$_{1+\delta}$Se does display nematic order~\cite{Watson15:91,Li20:10,Baek15:14} with a breaking of the C$_{4}$ tetragonal symmetry at 90 K~\cite{McQueen09:103} \textit{without} the formation of spatially long ranged~\cite{He18:97} magnetic order.   Despite many differences in the structural properties and magnetism in chalcogenides and pnictides, the electronics have been suggested to be quite similar with comparable Fermi surfaces,~\cite{Subedi08:78} and nematic orders~\cite{Pfau19:123} making the study of the dynamics driving this combined magneto-structural transition relevant.  In this paper, we investigate the soft acoustic phonon dynamics associated with this structural transition in bicollinear antiferromagnetically ordered Fe$_{1.057(7)}$Te finding a softening of the $C_{66}$ elastic constant on the THz timescale sampled with thermal neutron spectroscopy.

\begin{figure}[t]
	\begin{center}
		\includegraphics[]{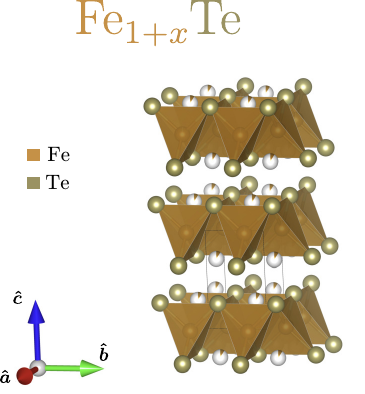}
	\end{center}
	\caption{High temperature tetragonal crystal structure of Fe$_{1+x}$Te: interstitial iron (in white and brown) located between weakly bonded layers of FeTe$_{4}$ tetrahedra. Figure made using VESTA~\cite{Momma11:44}.} 
	\label{fig:FT_struct}
\end{figure}

\begin{figure}[!]
	\begin{center}
		\includegraphics[width=\columnwidth]{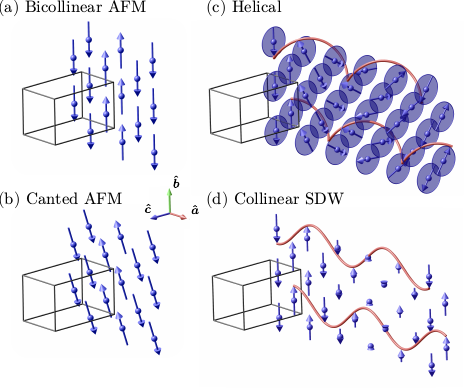}
	\end{center}
	\caption{Reported magnetic ground states for Fe$_{1+x}$Te: (a) bicollinear antiferromagnetism (AFM) found in low $x<0.12$ compounds, (b) canted AFM (in the $(bc)$-plane) measured on the surface layer of Fe$_{1+x}$Te samples, (c) helical incommensurate order found in high $x>0.12$ compounds, (d) collinear spin-density wave order at intermediate $x\approx0.12$ iron concentration. Only moments near $z=0$ are shown for clarity. Figures made using \textsc{Mag2Pol} \cite{Qureshi19:52}.}
	\label{fig:FT_mag}
\end{figure}

The structure of Fe$_{1+x}$Te and its anion substituted counterparts is deceptively simple, being based on single layers (Fig. \ref{fig:FT_struct}).  However, the physical properties are extremely sensitive to small amounts of interstitial iron that reside between the van der Waals layers as outlined in a number of studies.~\cite{Rodriguez11:84,Thampy12:108,Sun16:6,McQueen09:79,Vivanco16:242,Tang16:6,Bao09:102,Chen09:79,Ducatman14:90,Li21:20}  For small interstitial iron concentrations ($x \leq 0.12$), a high temperature tetragonal $P4/nmm$ structure (space group No. 129) undergoes a first-order transition to a $P2_{1}/m$ (space group No. 11) monoclinic unit cell with the magnetic moments carried by the iron sites ordering in a bicollinear structure at the same temperature as shown in Fig. \ref{fig:FT_mag} $(a)$.  Interestingly, scanning tunneling measurements find that near the physical crystal surface of a sample displaying this bicollinear magnetic structure, the spin arrangement is canted along the $c$-axis displayed schematically in Fig. \ref{fig:FT_mag} $(b)$.~\cite{Trainer21:103}  This magneto-structural transition also has implications on the electronic transport properties with resistivity displaying a transition from a semi (poor)-metallic response to a metallic one below the transition.~\cite{Rodriguez13:88}  This occurs at the same temperature as an energetic gap opens in the spin excitation spectrum, measured with neutrons. This temperature dependent spin-gap has been related to the precipitous drop in resistivity at the magneto-structural transition given the consequential removal of charge scattering channels at low-energies.  

For large concentrations of interstitial iron $x \geq 0.12$, the low temperature magnetic structure is helical (Fig. \ref{fig:FT_mag} $c$) and magnetic order also occurs at the same temperature as a $Pmmm$ (space group No. 59) orthorhombic unit cell replaces the high temperature tetragonal ($P4/nmm$).  Resistivity measurements do not show a dramatic change at the transition temperature displaying semi (poor) metallic responses at both high and low temperatures on either side of the magneto-structural transition.~\cite{Rodriguez13:88,Koz13:88}  The magnetic excitation spectrum remains energetically gapless, within experiment resolution dictated by neutron scattering, in both high and low temperature phases.  Analogous to the effects of interstitial iron on the resistivity in Fe$_{1+x}$Te, anion substituted and superconducting Fe$_{1+x}$Te$_{0.7}$Se$_{0.3}$ displays a gradual suppression of superconductivity, and the magnetic resonance peak in the neutron response, with increased interstitial iron doping.~\cite{Rodriguez11:2,Stock12:85}  We note that the magnetically ordered bicollinear and spiral phases are separated by a region near $x\sim 0.12$ where the magnetic structure is defined by a collinear spin-density wave phase~\cite{Rossler11:84,Zaliznyak12:85,Materne15:115} schematically illustrated in Fig. \ref{fig:FT_mag} $(d)$. 

In this paper, motivated by recent work investigating acoustic instabilities that drive the structural transitions from a tetragonal to an orthorhombic unit cell in the pnictides and corresponding intermediate nematic phases, we investigate the acoustic fluctuations corresponding to homogeneous deformations of the structural lattice in Fe$_{1.057(7)}$Te.  We choose this concentration as it corresponds to a magneto-structural transition to bicollinear order in the magnetism and, in terms of the electronics, is where the resistivity displays a concomitant transition from a semi (or poorly) metallic to metallic.  However, in contrast to the pnictides, this concentration is not superconducting and the low temperature structural unit cell is monoclinic, \textit{not} orthorhombic.  However, we do find a softening of an acoustic branch corresponding to the $C_{66}$ shear modulus.  We compare the results and speculate on the differences in terms of orbital magnetism in the conclusions and discussion.  

\section{Experimental details}

\textit{Materials:} The sample studied here was the exact same crystal of Fe$_{1.057(7)}$Te previously investigated by us using a number of scattering and transport techniques.  Structural studies using neutron diffraction found that this sample underwent a structural transition from a high temperature tetragonal to low temperature monoclinic unit cell at T$_{S}$ $\sim$ 70 K.~\cite{Rodriguez11:84}  Simultaneously, the formation of bulk bicollinear magnetic order occurs with the magnetic moments oriented in the $a-b$ plane.~\cite{Trainer21:103}  This magneto-structural transition also coincides with a transition from a semimetallic at high temperatures to metallic at low temperatures.~\cite{Rodriguez13:88}  Details on the single crystal growth conditions are described in Ref. \onlinecite{Stock11:84}.

\textit{X-ray Diffraction:}  To characterize the structural distortion in our powder sample and confirm it correlates with the magnetic transition, temperature dependent x-ray diffraction was performed using a Rigaku Smartlab diffractometer combined with a PheniX displex from Oxford Cryosystems.  This system allowed the measurement of powder diffraction patterns from plate-like samples using Bragg-Brentano geometry over temperatures ranging $T$=12-300 K.  An x-ray wavelength of $\lambda=1.54$ \AA\ was monochromated using a Johansson Monochromator.

\begin{figure}[t]
	\begin{center}
		\includegraphics[]{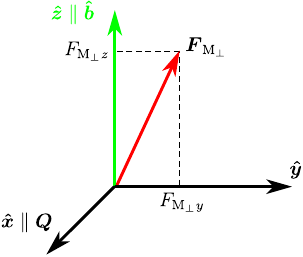}
	\end{center}
	\caption{Polarimetry local coordinates for Fe$_{1+x}$Te. The scattering vector $\bf{Q}$ lies in the $(H0L)$ scattering plane, and the vertical axis is along $\bf{\hat{b}}$.} 
	\label{fig:FT_local}
\end{figure}

\textit{Neutron Diffraction:}  Spherical neutron polarimetry experiments were performed using CRYOPAD~\cite{Tasset99:69,Geffray05:131} on the hot neutron diffractometer D3, at the ILL~\cite{FT2021D3} using a wavelength of $\lambda=0.85$ \AA\ selected by the (111) reflection of a Cu$_{2}$MnAl Heusler monochromator. The sample was a cylindrical single crystal of approximate height 4 mm and diameter 7 mm, mounted in the $(H0L)$ scattering plane cut from the same crystal used for neutron spectroscopy in this paper and previously by us and described above. The $b$-axis of the crystal (referenced in the high temperature tetragonal structure) is the vertical axis in the local laboratory coordinates of the diffractometer.  As shown in Fig. \ref{fig:FT_local}, magnetic scattering is only sensitive to the component of the magnetic structure factor perpendicular to the scattering vector following selection rules of magnetic neutron diffraction.  In the case of collinear magnetic structures like the bicollinear commensurate (Fig. \ref{fig:FT_mag} $a$) or the incommensurate antiferromagnetism (Fig. \ref{fig:FT_mag} $d$), the magnetic structure factor lies on the $z$-axis, so that $|{\bf{F}}_{\mathrm{M}_{\perp}}|=|F_{{\mathrm{M}}_{\perp z}}|$. Any presence of magnetic moments apart from the $b$-axis would result in a non-zero $F_{\mathrm{M}_{\perp y}}$ component (depending on the measured magnetic reflection), giving $|F_{\mathrm{M}_{\perp z}}|<|\bf{F}_{M_{\perp}}|$. This could be a sign of helical ordering (Fig. \ref{fig:FT_mag} $c$) or canted AFM (Fig. \ref{fig:FT_mag} $b$).

\begin{figure}[t]
	\begin{center}
		\includegraphics[width=78mm,trim=0.0cm 0.5cm 0cm 0cm,clip=true]{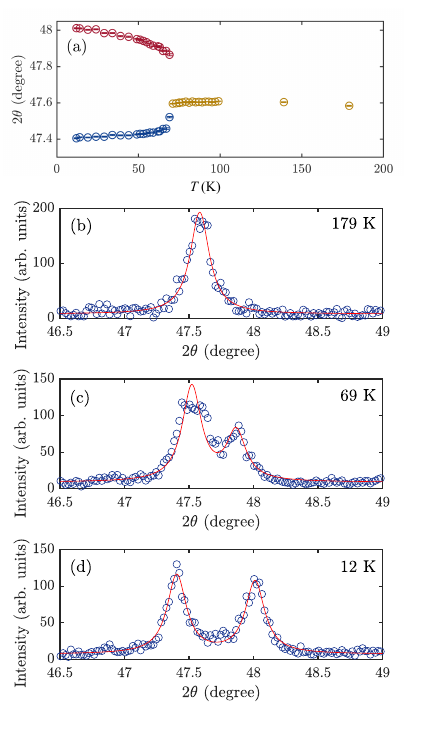}
	\end{center}
	\caption{Monochromatic X-ray diffraction (Rigaku Smartlab) data characterizing the structural transition through scanning the $\vec{Q}$=(200) Bragg peak. $(a)$ illustrates the temperature dependence of peaks from a high temperature tetragonal to a low temperature monoclinic phase- evidenced by the peak splitting at the transition temperature.  $(b-d)$ shows the representative peaks fitted in both high temperature and low temperature structural phases.} 
	\label{fig:structural}
\end{figure}

\begin{figure}[t]
	\begin{center}
		\includegraphics[width=78mm,trim=0.5cm 0.7cm 0cm 0cm,clip=true]{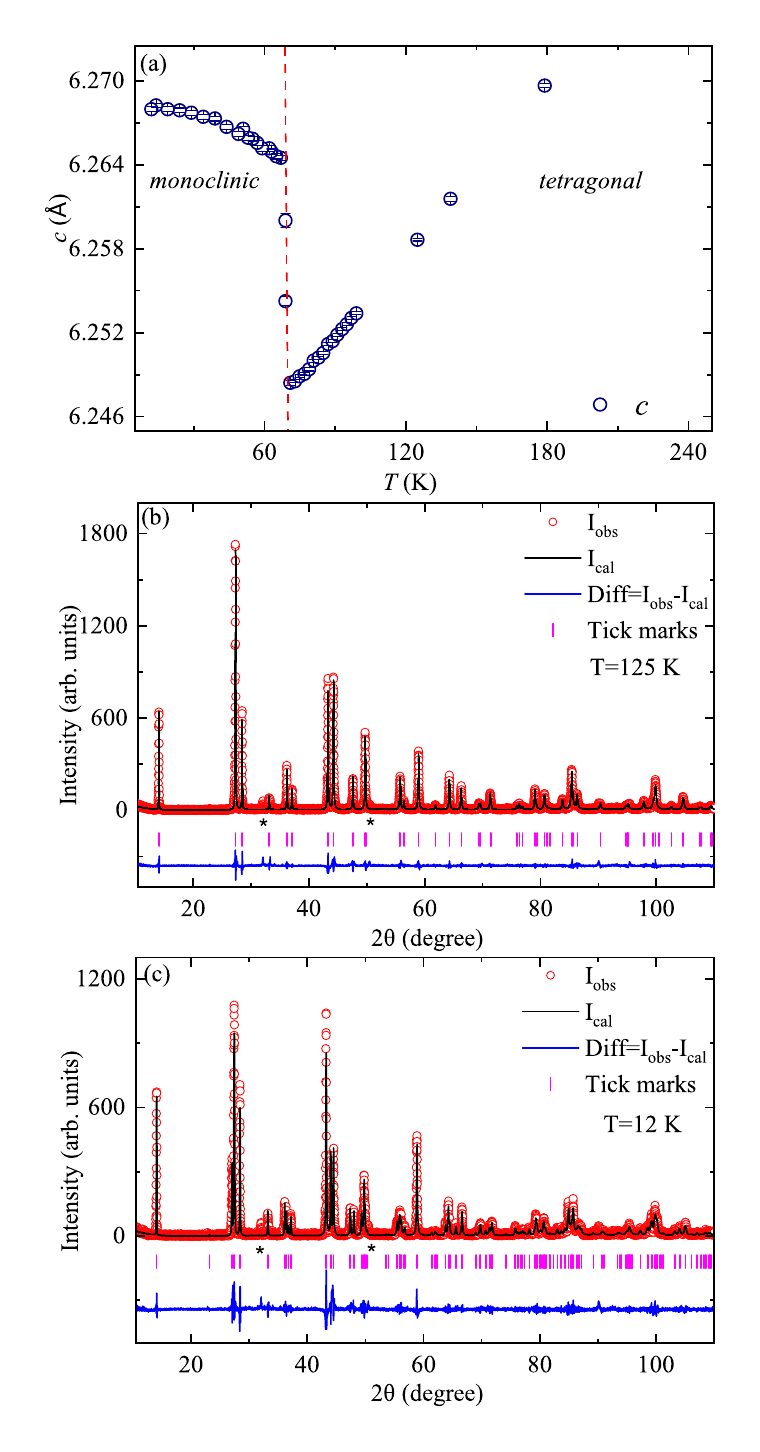}
	\end{center}
	\caption{Monochromatic x-ray diffraction (Rigaku Smartlab) data taken on a ground powdered piece from the single crystal used for neutron spectroscopy. Panel $(a)$ illustrates the $c$-lattice constant displaying a sharp first-order transition from a high temperature tetragonal unit cell to a low temperature monoclinic (data taken on warming).  $(b-c)$ displays representative x-ray diffraction data in both high temperature and low temperature structural phases. The asterisk corresponds to impurity FeTe$_{2}$ ($\sim$ 32$^{\circ}$) and Fe ($\sim$ 50.5$^{\circ}$) phases.} 
	\label{fig:FT_Diff}
\end{figure}

\textit{Neutron Spectroscopy:}  Neutron spectroscopy measurements to investigate acoustic lattice instabilities were performed on the EIGER thermal triple-axis spectrometer (PSI, Switzerland).~\cite{Stuhr16:853}  The Fe$_{1.057(7)}$Te single crystal was oriented such that Bragg reflections of the form $(HK0)$ lay within the horizontal scattering plane. For all measurements, the final energy was fixed to $E_{f}$=14.7 meV using a pyrolytic graphite (002) horizontally focussed analyzer crystal, and the incident energy was varied using a vertically and horizontally focused graphite (002) monochromator.  This spectrometer configuration defined the energy transfer as $E=E_{i}-E_{f}$.  Higher harmonics present in the neutron beam were suppressed using a graphite filter on the scattered side.  Counting times were determined by an incident beam monitor with a low counting efficiency and were corrected for higher harmonic contamination of this detector from the monochromator.~\cite{Shirane:book}  

\section{Results}

In this section, we outline the experimental results.  Given the complexity of the static magnetism in Fe$_{1+x}$Te and the presence of a collinear spin-density wave phase, for interstitial iron concentrations of $x\sim 0.12$, it is important to establish the magnetism near the critical temperature in our single crystal and in particular if there is a presence of collinear spin density wave order that differs from the low temperature bicollinear magnetic order.  We first study the structural transition using temperature dependent x-ray diffraction and then apply spherical neutron polarimetry to investigate the static magnetism near the magneto-structural transition.  The combination of this structural x-ray study and magnetic neutron investigations shows a magneto-structural transition from a high temperature paramagnetic tetragonal phase to a low-temperature monoclinc bicollinear phase.  We do not observe evidence for an intermediate spin density wave phase (as reported in larger iron concentrations) nor an intermediate phase where the tetragonal C$_{4}$ symmetry is broken while time reversal is not.  This fixes our single crystal in the Fe$_{1+x}$Te phase diagram within the magnetic bicollinear ordered phase and not where static incommensurate or spin-density wave order is reported in concentrations near $x\sim 0.12$.  After characterizing the magneto-structural transition, we then apply neutron spectroscopy to observe a softening and recovery of the acoustic phonon branch sensitive to the $C_{66}$ shear modulus.

\subsection{Structural distortion}

We first analyze the structural distortion in Fe$_{1.057(7)}$Te using powder x-ray diffraction off a finely ground powder taken from a piece of our single crystal used for neutron spectroscopy discussed below.  In Fig. \ref{fig:structural}, we track the temperature dependence of the (200) nuclear Bragg peak with peak positions plotted in Fig. \ref{fig:structural} $(a)$ based on fits to a double lorentzian with illustrative fits shown in Fig. \ref{fig:structural} $(b-d)$.  At high temperatures above T$_{S}\sim$ 70 K, a single peak is observed as expected based on a tetragonal unit cell.  For temperatures well below 70 K, this peak splits indicating that the $a$ and $b$ lattice constants are no longer equivalent confirming a breaking of the high temperature C$_{4}$ structural tetragonal symmetry. However, no additional peak splitting was observed at any temperatures down to 12 K, within our instrumental resolution.

This structural transition is further analyzed in Fig. \ref{fig:FT_Diff} which shows the temperature dependent $c$-lattice constant (Fig. \ref{fig:FT_Diff} $a$) and representative profile refinements (Le Bail) in Fig. \ref{fig:FT_Diff} $(b,c)$ that fit the unit cell shape.  Temperature dependent profile refinements of the diffraction data were performed using TOPAS~\cite{TOPAS} and Jana software~\cite{Petricek14:229} with good fits found for a high temperature $P4/nmm$ with a tetragonal unit cell (space group No. 129) and a low temperature with a $P2_{1}/m$ (space group No. 11) monoclinic unit cell.  Both the temperature dependence of the $c$-lattice constant\
(Fig. \ref{fig:FT_Diff} $a$) and the splitting of the tetragonal (200) Bragg peak (Fig. \ref{fig:structural} $a$) are indicative of a discontinuous first-order structural transition at $\sim$ 70 K.

\subsection{Spherical neutron polarimetry}

\begin{figure}[t]
	\begin{center}
		\includegraphics[]{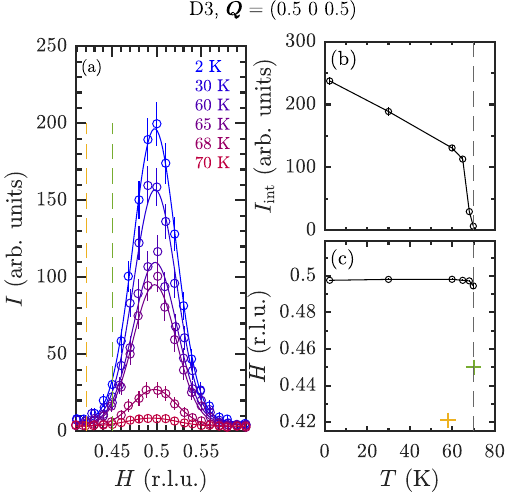}
	\end{center}
	\caption{(a) $H$-scans along $(H,0,0.5)$ for different temperatures, fitted to Gaussians (continuous lines).  Temperature dependence of (b) the integrated intensities, (c) the center of the magnetic reflection. $T_{N}\sim$ 70 K is indicated in dashed gray lines. The reported values $\delta=0.421(1)$ at 57 K \cite{Rodriguez11:84} and $\delta=0.45$ at 70 K \cite{Stock11:84} are shown in yellow and green in (a) dashed lines, (c) crosses.} 
	\label{fig:FT_Qscan}
\end{figure}

Having confirmed the structural first-order transition with powder x-rays, we now investigate the magnetic response through this transition using polarized neutrons.  To determine the possibility of the presence of incommensurate or spin density wave fluctuations that complicate the interpretation of the magneto-structural transition, there are two points that need to be considered.  First, whether the wave vector characterizing the magnetism is incommensurate and hence not coincident with the commensurate $H={1\over 2}$ position, and second if there is an observable canting of the magnetic moments as suggested that may exist from scanning tunneling spectroscopy measurements.~\cite{Trainer21:103,Trainer19:5}  

\renewcommand{\arraystretch}{1.15}
\begin{table*}
	\caption{Polarization matrices measured on D3 for different reflections at different temperatures.}
	\label{tab:FT_polar}
	\begin{tabular}{c c c c }
		\hline\hline
		$\vec{Q}$ & $T=2 K$ &   $T=30 K$ & $T=60 K$\\ 
		\hline
		$(\frac{1}{2}\ 0\ \frac{1}{2})$ & 
		$	
		\begin{pmatrix}-0.985(4) & 0.036(7) & 0.099(7) \\
			-0.002(7) & -0.983(4) & 0.093(7) \\
			0.111(7) & 0.070(7) & 0.981(4) 
		\end{pmatrix}
		$ & $ 
		\begin{pmatrix}-0.993(5) & 0.029(8) & 0.080(8) \\
			0.004(8) & -0.980(5) & 0.090(8) \\
			0.117(8) & 0.069(8) & 0.983(5) 
		\end{pmatrix}
		$ & $
		\begin{pmatrix}-0.976(7) & 0.02(1) & 0.10(1) \\
			0.00(1) & -0.983(7) & 0.07(1) \\
			0.10(1) & 0.09(1) & 0.979(7) 
		\end{pmatrix}
		$  		
		\\
		$(\frac{3}{2}\ 0\ \frac{1}{2})$ & 
		$\begin{pmatrix}-0.97(2) & 0.02(3) & 0.10(3) \\
			0.02(3) & -0.93(2) & 0.02(3) \\
			0.13(3) & 0.03(3) & 1.01(2) 
		\end{pmatrix}$ &
		$
		\begin{pmatrix}-1.01(2) & 0.04(3) & 0.13(3) \\
			-0.00(3) & -1.00(3) & 0.05(3) \\
			0.14(3) & 0.07(3) & 1.03(3) 
		\end{pmatrix}
		$ &
		$
		\begin{pmatrix}-0.98(3) & 0.06(4) & 0.08(4) \\
			0.06(4) & -1.00(3) & 0.03(3) \\
			0.16(4) & 0.11(4) & 0.97(3) 
		\end{pmatrix}
		$
		
		\\
		$(\frac{3}{2}\ 0\ \frac{3}{2})$ &
		$\begin{pmatrix}-0.97(3) & 0.06(4) & 0.14(4) \\
			0.02(4) & -0.96(3) & 0.13(4) \\
			0.12(4) & 0.04(4) & 0.98(3) 
		\end{pmatrix}$ &
		$
		\begin{pmatrix}-0.99(3) & 0.06(4) & 0.00(4) \\
			0.02(4) & -1.06(4) & 0.16(4) \\
			0.03(4) & 0.05(4) & 0.97(3) 
		\end{pmatrix}$&
		$
		\begin{pmatrix}-1.06(5) & -0.08(5) & 0.04(5) \\
			0.05(5) & -0.92(5) & 0.11(5) \\
			0.08(6) & 0.25(6) & 0.90(5) 
		\end{pmatrix}
		$
	\end{tabular}
	\begin{tabular}{c c c c}
		\hline \hline
		$\vec{Q}$  &   $T=65 K$ & $T=68 K$ &\\ 
		\hline
		$(\frac{1}{2}\ 0\ \frac{1}{2})
		$ & $	    
		\begin{pmatrix}-0.973(7) & 0.045(9) & 0.097(9) \\
			0.004(9) & -0.976(7) & 0.092(9) \\
			0.122(9) & 0.095(9) & 0.984(7) 
		\end{pmatrix}
		$ &$
		\begin{pmatrix}-0.93(2) & 0.04(2) & 0.11(2) \\
			0.01(2) & -0.93(2) & 0.09(2) \\
			0.10(2) & 0.08(2) & 0.98(2) 
		\end{pmatrix}
		$	     
		\\
		$(\frac{3}{2}\ 0\ \frac{1}{2})$ & 
		$
		\begin{pmatrix}-0.97(3) & 0.02(3) & 0.11(3) \\
			0.03(3) & -0.96(3) & -0.00(3) \\
			0.11(4) & 0.08(3) & 0.96(3) 
		\end{pmatrix}
		$&$
		\begin{pmatrix}-1.00(5) & 0.01(6) & 0.15(5) \\
			0.05(5) & -0.80(6) & -0.06(6) \\
			0.09(6) & 0.05(5) & 0.93(6) 
		\end{pmatrix}
		$   	
		\\
		$(\frac{3}{2}\ 0\ \frac{3}{2})$ & $
		\begin{pmatrix}-0.98(3) & 0.03(4) & 0.12(4) \\
			0.01(4) & -0.98(4) & 0.06(4) \\
			0.11(4) & 0.04(4) & 0.93(4) 
		\end{pmatrix}
		$
		\\
		$(2\ 0\ 0)$ & &
		$\begin{pmatrix}0.994(6) & -0.006(6) & -0.030(7) \\
			-0.044(7) & 0.999(6) & -0.005(7) \\
			-0.007(7) & 0.032(7) & 0.999(5) 
		\end{pmatrix}$ \\
		\hline	\hline
	\end{tabular}		
\end{table*}

To first determine whether the static (measured at the elastic $E$=0 energy position) magnetic order is commensurate or incommensurate close to the N\'eel temperature, scans along $(H,0,0.5)$ were measured as a function of temperature, as shown in Fig. \ref{fig:FT_Qscan} $(a)$.  The N\'eel temperature defining the onset of bicollinear magnetic order was found to be at $T_{S}\sim$ 70 K (Fig. \ref{fig:FT_Qscan} $b$) with the sharp onset of intensity characterized by the first-order nature of the transition. The ordering wave vector is along the [1 0 0] direction as plotted in Fig. \ref{fig:FT_Qscan} $(c)$ and illustrates an almost immediate onset of commensurate (within experimental error) H=0.5 magnetism at the magneto-structural transition of $T_{S} \sim$ 70 K.  These results contrast with the evolution from incommensurate $\mathrm{k}=(\delta,0,\frac{1}{2})$ to commensurate $\mathrm{k}=(\frac{1}{2},0,\frac{1}{2})$ reported in Refs.~\onlinecite{Rodriguez11:84} ($\delta=0.421(1)$ at 57 K for Fe$_{1.09(1)}$Te). Similar incommensurate wave vectors have been reported with copper substitution.~\cite{Valdivia15:91}  We note that an incommensurate wave vector of $\delta=0.45$ was reported in the magnetic dynamics at low-energy transfers near T=70 K in this exact same crystal of Fe$_{1.057(7)}$Te and polarization of the fluctuations measured with polarized neutrons to be anisotropic.~\cite{Stock17:95}  At temperatures in the paramagnetic phase, these fluctuations are observed to continuously evolve from the incommensurate wavevector to the commensurate position at the elastic line (see Figure 2 in Ref. \onlinecite{Stock17:95}).  Returning to the static properties measured at the elastic (E=0) position, the incommensurate ordering wave vectors of $\delta$=0.42 or 0.45 r.l.u. reported for other Fe$_{1+x}$Te compounds with low temperature bicollinear order are indicated in the $(H,0,0.5)$ scans in Fig. \ref{fig:FT_Qscan} $(a)$ and are not consistent with the wave vector scans that characterize the magnetic order in Fe$_{1.057(7)}$Te at any of the measured temperatures.  Indeed, we can rule out peaks at these positions with an intensity less than 2\% of the commensurate peak measured at $H$=0.5.  This indicates that within the experimental resolution that the magneto-structural transition at $\sim$70 K is to commensurate bicollinear order with no observable intermediate incommensurate magnetic static scattering as in larger interstitial iron concentrations.

The establishment of a commensurate ordering wave vector at all observable temperatures indicates a lack of a static moment modulated spin density wave phase or spiral helical order.  We now analyze the polarization matrix elements and their temperature dependence. 

We first discuss the matrix elements of the polarization matrix.  For a pure elastic magnetic reflection, in the absence of any chiral terms, the polarization matrix for a fully polarized incident beam is given by,

\begin{equation} \label{Eqn:polmat}
	\mathsf{P} = \begin{pmatrix}
		-1 & 0 & 0 \\
		0 & {{|F_{\mathrm{M}_{\perp y}}|^2 - |F_{\mathrm{M}_{\perp z}}|^2} \over {|F_{\mathrm{M}_{\perp}}|^2}} & {{2 \mathcal{R}e \{ F_{\mathrm{M}_{\perp y}} F_{\mathrm{M}_{\perp z}}^*\}}\over {|F_{\mathrm{M}_{\perp}}|^2}} \\
		0 & {{2 \mathcal{R}e \{ F_{\mathrm{M}_{\perp z}} F_{\mathrm{M}_{\perp y}}^*\}}\over {|F_{\mathrm{M}_{\perp}}|^2}}  &  {{|F_{\mathrm{M}_{\perp z}}|^2 - |F_{\mathrm{M}_{\perp y}}|^2} \over {|F_{\mathrm{M}_{\perp}}|^2}}	
	\end{pmatrix}.
\end{equation}

\noindent We note that chiral terms would produce non zero and equal $\mathsf{P}_{yx}$ and $\mathsf{P}_{zx}$ matrix elements (see Ref. \onlinecite{Giles-Donovan20:102} Eqn. 24).  The matrix elements $\mathsf{P}_{xy}$ and $\mathsf{P}_{xz}$ are identically equal to 0 for purely magnetic scattering.  

As illustrated in Fig. \ref{fig:FT_local}, magnetic moments fully aligned along the $b$-axis would only give a $F_{\mathrm{M}_{\perp z}}$ component. Any canting from the $b$-axis would result in a non-zero $F_{\mathrm{M}_{\perp y}}$ component, and the reduction of the amplitude of the diagonal elements $P_{yy}=-P_{zz}$ from 1. In addition, the structural transition from tetragonal to monoclinic symmetry in Fe$_{1+x}$Te would lead to four structural domains. In the case of equi-populated domains, the off-diagonal elements $\mathsf{P}_{yz}$ and $\mathsf{P}_{zy}$ would average to zero (as discussed in Ref. \onlinecite{Trainer21:103}), leading to a purely diagonal polarization matrix

\begin{equation}\label{eq:diago}
	\mathsf{P} (i \rightarrow f)=\begin{pmatrix} 
		-1 & 0 & 0\\ 
		0 & -x & 0\\
		0 & 0 & x
	\end{pmatrix},
\end{equation}

\noindent where $x=1$ if the moments are along $\hat{b}$ and $|x|<1$ if they are canted.  

The polarization matrices for three magnetic reflections $(\frac{1}{2}\ 0\ \frac{1}{2})$, $(\frac{3}{2}\ 0\ \frac{1}{2})$ and $(\frac{3}{2}\ 0\ \frac{3}{2})$ were measured at five different temperatures and listed in Table \ref{tab:FT_polar}. Each polarization matrix was corrected for the initial polarization on D3 with $p_0=0.935$. The decay of the $^3$He spin filter efficiency was tracked by measuring the $\mathsf{P}_{zz}$ matrix element of the nuclear Bragg peak $(2\ 0\ 0)$ (where magnetic scattering is absent) at regular time intervals. We note that $\mathsf{P}_{zz}\equiv 1$ for a nuclear peak and perfect initial and final neutron beam polarization. All polarization matrices listed Table \ref{tab:FT_polar} were corrected for this change in spin filter efficiency.

\begin{figure}[t]
	\begin{center}
		\includegraphics[]{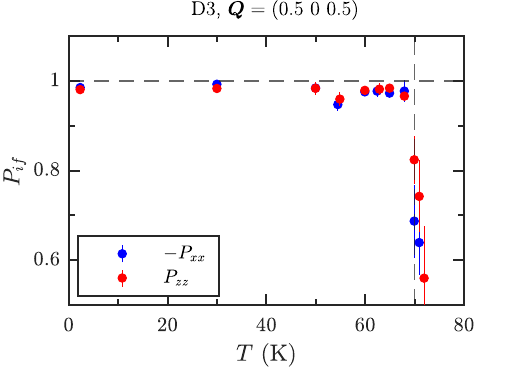}
	\end{center}
	\caption{Temperature dependence of the polarization matrix components $\mathsf{P}_{xx}$ and $\mathsf{P}_{zz}$ for the magnetic reflection at $\vec{Q}=(0.5\ 0\ 0.5)$.} 
	\label{fig:FT_Pzz}
\end{figure}

Finally, all the polarization matrices in Table \ref{tab:FT_polar} follow a diagonal form as represented by Eqn. \ref{eq:diago}, where $x=1$. We note that some off-diagonal terms are non-zero within the calculated statistical uncertainties based solely on Poisson counting statistics. This is particularly true for $\mathsf{P}_{xz}$ and $\mathsf{P}_{xy}$ which should vanish for a pure magnetic reflection. Non-zero values for $\mathsf{P}_{zx}$ are also found, but they do not correspond to a chiral term which would necessitate $\mathsf{P}_{yx}\equiv \mathsf{P}_{zx}$.  As discussed in Ref. \onlinecite{Giles-Donovan20:102}, as well as the counting statistical errors there are other systematic errors in the use of cryopad which are much larger.  In particular, misalignment of the sample with respect to the neutron beam can increase the errorbars on the matrix significantly.  The precision of the polarization direction of the incident beam in relation to the crystallographic axes is $\sim 2^\circ$.~\cite{Heil99:267}  This errors are more obvious for the off diagonal components of the weaker magnetic peaks over the intense nuclear peaks as the integrated intensities are more susceptible to background measurements and relative subtractions.  Also, the sample was aligned at high temperatures in the tetragonal phase, and on entering the monoclinic phase may introduce similar misalignments of the sample with respect to the neutron beam.  Such alignment errors are not observable given the diffractometer resolution and sample mosaic.  

Based on this, we conclude that all of the measured matrices at all temperatures are diagonal within experimental error, following the form of Eqn. \ref{eq:diago}.  With deviations of the absolute value of the diagonal elements from 1 being sensitive to canting or helical ordering, we track the temperature dependence of these matrix elements in Fig. \ref{fig:FT_Pzz}.  This figure shows an abrupt onset of these matrix elements at the first-order magneto-structural transition at 70 K with the errorbars only representing Poisson counting statistics. The matrix elements calculated above the transition are unphysical, as the magnetic intensity vanishes as shown in Fig. \ref{fig:FT_Qscan} $(b)$. These elements rather correspond to the variation of background intensities in the different polarization channels. Based on the discussion above these values are consistent with 1 and show no observable significant change with temperature.  Based on the temperature dependence and the discussion of the matrix elements, we conclude that Fe$_{1.057(7)}$Te undergoes a first-order phase transition to a bicollinear magnetic phase at low temperatures.  We do not find evidence of any helical or incommensurate spin-density wave phases reported for larger iron concentrations of Fe$_{1+x}$Te near the magnetic phase transition.

\subsection{Neutron spectroscopy of acoustic phonons}

\begin{figure}[t]
	\begin{center}
		\includegraphics[width=88mm,trim=0.5cm 1.8cm 0cm 0cm,clip=true]{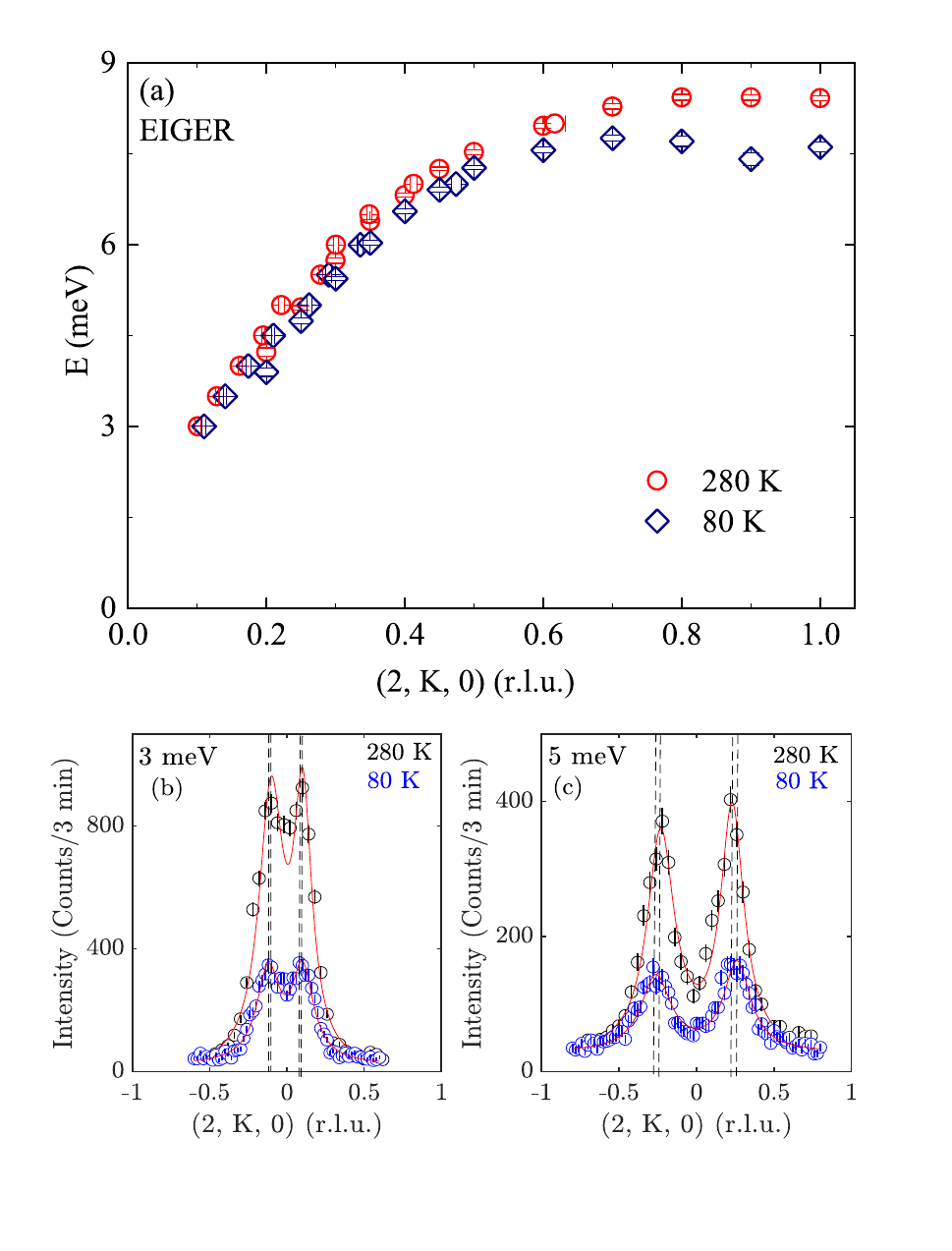}
	\end{center}
	\caption{$(a)$ The temperature dependence of the measured dispersion.  $(b-c)$ Illustrative constant $E$-scans at 3 meV and 5 meV, respectively. The dashed lines represent the fitted temperature dependent shift in the $Q$-position.} 
	\label{fig:ph_disp}
\end{figure}

Having discussed the static properties of the magneto-structural transition, we now present the lattice dynamics near the structural transition with the goal of identifying the soft lattice dynamics associated with the magneto-structural transition in Fe$_{1.057(7)}$Te.  We note the magnetic dynamics near this phase transition are discussed in Refs. \onlinecite{Stock17:95,Parshall12:85}.  Motivated by recent work on the pnictides studying soft elastic constants tied with the formation of nematic order,~\cite{Merrit20:124,Kauth20:14,Wu21:126} we investigate the low-energy acoustic phonons in our crystal of Fe$_{1.057(7)}$Te linked with uniform deformations of the lattice.

We focus our analysis by studying acoustic phonons that are polarized along [100] and propagating along [010] as sampled by performing scans near $\vec{Q}=(2, K, 0)$.   As discussed in Ref. \onlinecite{Cowley76:13}, in a tetragonal phase these acoustic phonons are sensitive to the $C_{66}$ elastic constant as in limit of $q \rightarrow 0$ the slope of the acoustic phonon dispersion is directly proportional to the velocity $c$ by $E=\hbar c q$.  The acoustic phonon velocity is sensitive to the elastic constant through $c = \sqrt{C_{66}/\rho}$, where $\rho$ is the density of Fe$_{1.057(7)}$Te.~\cite{Dove:book}  

The measured neutron scattering intensity at a particular energy transfer defined $\hbar\omega \equiv E\equiv E_{i}-E_{f}$ on a triple-axis instrument at a momentum transfer $\vec{Q}\equiv \vec{k}_{i}-\vec{k}_{f}$, with a monitor detector before the sample, is directly proportional to the structure factor $S(\vec{Q},E)$.  This in turn is related to the imaginary part of the susceptibility $\chi''(Q,\omega)$ by,

\begin{equation}\label{eq:def}
	I(\vec{Q},E) \propto S(\vec{Q},E) \equiv {1 \over \pi} [n(E)+1] \chi''(\vec{Q},E).
\end{equation} 

\noindent In this experiment we have modeled the acoustic phonon scattering in terms of an antisymmetric sum of Lorentzians,

\begin{equation}\label{eq:SHO}
	\chi '' (\vec{Q}, E) \propto \left({1 \over {1+\left({{\omega-\Omega_{0}(\vec{Q})} \over \Gamma}\right)^{2}}}- {1 \over {1+\left({{\omega+\Omega_{0}(\vec{Q})} \over \Gamma}\right)^{2}}}\right) 
\end{equation}

\noindent where, $\hbar \Omega_{0}= \hbar c q$, and $\Gamma$ defines the energy linewidth which is inversely proportional to the acoustic phonon lifetime $\propto 1/\tau$.  The two Lorentzians are needed to ensure that $\chi''$ is an odd function, required by the principle of detailed balance applied to neutron spectroscopy.  In the discussion that follows, we have fit all constant momentum (energy) scans with this antisymmetric sum of two Lorentzians convolved with the energy resolution of the spectrometer. 

\begin{figure}[t]
	\begin{center}
		\includegraphics[width=97mm, trim=1.55cm 1.7cm 0cm 0.7cm,clip=true]{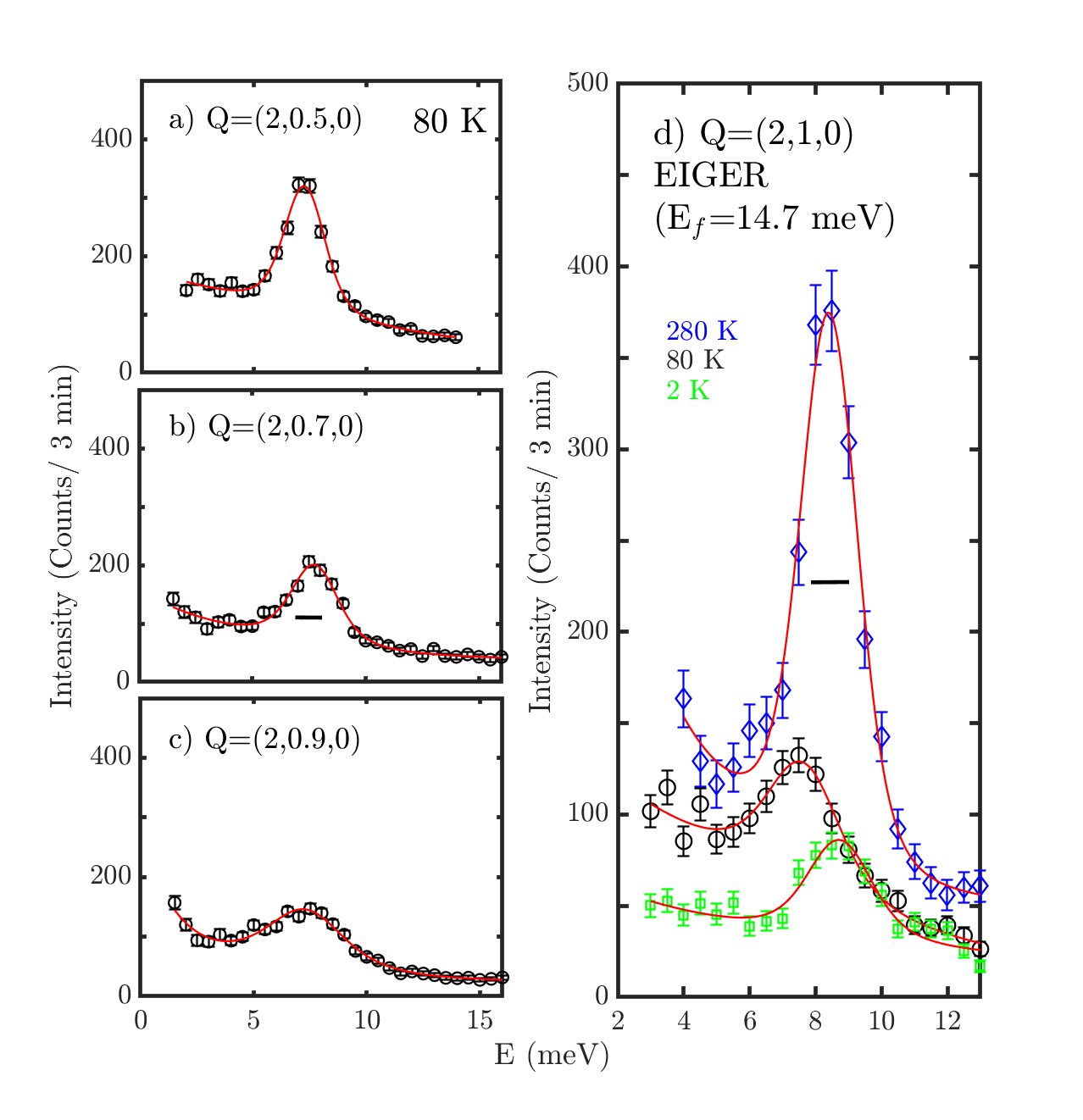}
	\end{center}
	\caption{$(a-c)$ Constant momentum scans sensitive to the elastic constant $C_{66}$ at 80 K illustrating dispersive acoustic lattice fluctuations measured using the EIGER triple-axis spectrometer.  $(d)$ Constant momentum scans at $\vec{Q}$=(2,1,0) illustrating a temperature dependence of this acoustic branch which softens at temperatures near the magnetostructural transition and then recovers at low temperatures.}   
	\label{fig:ph_constq}
\end{figure}

In Fig. \ref{fig:ph_disp} $(a)$, we plot the acoustic phonon dispersion throughout the Brillouin zone along $\vec{Q}=(2, K, 0)$ at both high temperature (280 K) and at 80 K, near the magneto-structural transition at $\sim$ 70 K constructed from a series of constant energy and momentum scans.  These measurements are done at considerable distance away from the Brillouin zone center with measurements starting at $q \sim 0.1$.  This results in the acoustic phonon dispersion not being in the purely linear regime where $E=\hbar c q$ which would require higher resolution techniques such as ultrasound hence not providing a good estimate of the elastic constants.  Representative constant energy scans are shown in Fig. \ref{fig:ph_disp} $(b,c)$ and are fit to two symmetrically displaced Lorentzians with differing linewidths to account for the different energy resolutions from tilting of the resolution ellipsoid (an effect known as focussing~\cite{Shirane:book}).  For the remainder of this section where we track the energy positions and linewidths, we apply scans on the focused side where the energy resolution is narrower.  Further representative constant momentum scans are illustrated in Fig. \ref{fig:ph_constq}.

Fig. \ref{fig:ph_disp} $(a)$ displays a clear softening of this acoustic phonon across the entire Brillouin zone with decreasing temperature with the most pronounced softening at the zone boundary $\vec{Q}$=(2,1,0).  This contrasts with the normal response of acoustic phonons which continuously harden in energy with decreasing temperature (see, for example, Fig. 14 of Ref. \onlinecite{Stock12:86}).  This is further reflected in the constant energy scans displayed in Fig. \ref{fig:ph_disp} $(b-c)$ which show a shift towards larger momentum indicating a lower energy in the acoustic phonon dispersion on comparing T=80 K and 280 K data.  We note that the acoustic phonon energy softens to a smaller extent in the limit $q \rightarrow 0$ and larger as $q$ approaches the Brillouin zone boundary at $\vec{Q}=(2,1,0)$.   Unlike some compounds where the acoustic phonon softens over a particular momentum range (for example in Ref. \onlinecite{Weber17:96}), we observe a softening across the entire Brillouin zone with the amount of softening in proportion to the wavevector $q$.  This indicates a softening of the elastic constant $C_{66}$.  

\begin{figure}[t]
	\begin{center}
		\includegraphics[width=95mm,trim=0.8cm 0.9cm 0cm 0.6cm,clip=true]{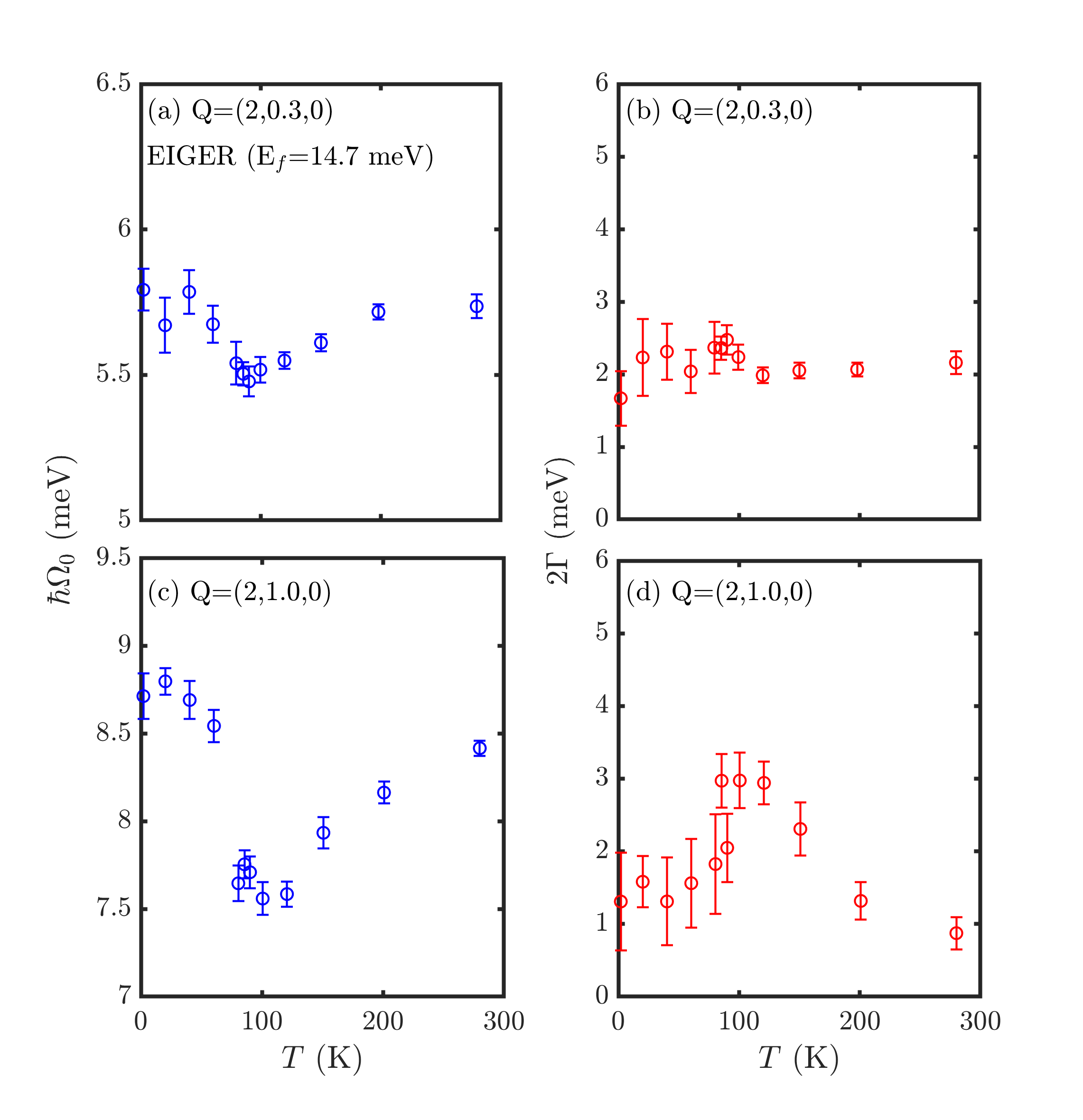}
	\end{center}
	\caption{The energy and full widths of the acoustic phonons measured at $(a-b)$ $\vec{Q}$=(2,0.3,0) and $(c-d)$ $\vec{Q}$=(2,1.0,0)} 
	\label{fig:ph_param}
\end{figure}

Given the softening is most evident at the zone boundary, we plot the energy position and lifetime at (2,1,0) and also (2, 0.3, 0) as a function of temperature in Fig. \ref{fig:ph_param}.  These parameters originate from fits to the above antisymmetrical lineshape with representative constant momentum scans throughout the Brillouin zone, and fits, shown in Fig. \ref{fig:ph_constq} $(a-d)$. Fig. \ref{fig:ph_constq} $(d)$ displays a clear softening of the acoustic phonon at the zone boundary $\vec{Q}=(2, 1, 0)$ near the magneto-structural transition with a similar and less pronounced softening at small momentum transfers $\vec{Q}=(2, 0.3, 0)$ (Fig. \ref{fig:ph_param} $a$).  We note that the softening is not complete and the energies only soften $\sim 10 \%$ of their values owing to the fact the magneto-structural transition in Fe$_{1.057(7)}$Te is first order.  This softening at the magneto-structural transition of the phonon energy is also accompanied by an increase in the acoustic phonon linewidths in energy, which is inversely proportional to the lifetime.  While constant momentum scans near $\vec{Q}=(2,0.3,0)$ (Fig. \ref{fig:ph_param} $b$) show little experimentally observable change in the linewidth near the magneto-structural transition, at the zone boundary $\vec{Q}=(2,1.0,0)$ (Fig. \ref{fig:ph_param}), a significant change in the phonon linewidth is observed near the magneto-structural transition.  Acoustic phonon measurements done with neutron spectroscopy observe an acoustic phonon softening and dampening of the acoustic phonon near the zone boundary.  

\section{Discussion and Conclusions}

To summarize our results, we have first characterized the static structural and magnetic properties of a Fe$_{1.057(7)}$Te single crystal using x-ray diffraction and spherical neutron polarimetry. These studies found a first-order magneto-structural transition from a tetragonal paramagnet to a bicollinear magnetic phase with a monoclinic unit cell.  We find no observable evidence of a static collinear spin-density wave near the magneto-structural transition temperature as is reported for larger interstitial iron concentrations and also observed in the dynamic response in this same Fe$_{1.057(7)}$Te single crystal.  Motivated by work on the pnictides studying soft acoustic fluctuations important for nematic correlations, we examined the temperature dependence of the acoustic phonons related to uniform distortions of the lattice.  We report a softening of the acoustic phonon branch dependent on the $C_{66}$ elastic constant and shear modulus.  This softening is not complete presumably due to the first-order nature of the magneto-structural transition, nor as large as in the pnictides where nematic order is present.  This softening is largest near the structural transition and coincident with the magnetic transition from a high temperature paramagnet to low temperature bicollinear order.  

While there have been several reports of optic phonon anomalies indicative of spin-phonon coupling in the chalcogenides~\cite{Homes16:93,Baum18:97,Popovic14:193}, our work is motivated by previous studies that have found acoustic phonons being the soft modes of orbitally driven phase transitions.~\cite{Gehring75:38,Weber17:96,Birgeneau74:10,Pytte71:3,Pytte73:8,Pavlovskiy18:97} We now discuss this magneto-structural transition in terms of a picture of localized magnetic moments as presented to understand the magnetism in this series of related chalcogenide compounds.~\cite{Turner09:80,Chen13:88,Lanata13:87}  While Fe$_{1+x}$Te displays an itinerant character with optical data on the pnictide compounds providing evidence that itinerant effects drive the spin-density wave phases,~\cite{Hu08:101} theoretical studies~\cite{Raghu08:77,Graser09:11,Daghofer10:81} have highlighted the importance of the localized $3d$ orbitals on the electronic properties and potential superconducting pairing.  In the context of nematic fluctuations, recent x-ray dichroism~\cite{Occhialini23} combined with strain measurements have found a strong orbital response at the onset of nematic order in FeSe and which has been previously supported by NMR results~\cite{Baek15:14}.  A lifting of the $d$ orbital degeneracy at the nematic transition has also been reported using photoemission.~\cite{Shimojima14:90}  These studies point to growing evidence that orbital physics is central to the formation of a nematic phase.

The localized magnetism in Fe$_{1+x}$Te is based upon Fe$^{2+}$ in a tetrahedral coordinated crystalline electric field environment which splits the 5 degenerate $d$-orbitals into lower doubly degenerate $|e_{g}\rangle$ and an upper triply degenerate $|t_{g}\rangle$ manifolds.~\cite{Abragam:book}  The 6 $d$ electrons of an Fe$^{2+}$ ion occupy these levels based on the Pauli exclusion principle and Hund's rules.  In this scenario, there is a choice on how to populate the electrons amongst the $|e_{g}\rangle$ and $|t_{g}\rangle$ manifolds with two energy scales defined by the Hund's coupling and the crystal field splitting which defines the energy gap between the low energy $|e_{g}\rangle$ and $|t_{g}\rangle$ free ion orbitals in a tetrahedral crystal field.  In the case that Hund's coupling is the dominant energy scale (as is the case in the weak and intermediate crystal field limit), then all 5 orbitals would be populated based on the Pauli principles and then double occupancy based on the energetics of the orbitals.  In the case that the crystal field energy is large (strong crystal field limit) the lower $|e_{g}\rangle$ are filled first and then followed by the upper $|t_{g}\rangle$ manifold once these are filled. These two limits result in very different magnetic ground states.  For the weak or intermediate crystal field limit, an orbital singlet ground state occurs with a net spin of $S=2$. For the strong crystal field limit, an orbital triplet ground state occurs with $S=1$.

The static magnetism in Fe$_{1+x}$Te has been investigated for a number of interstitial iron concentrations with neutron diffraction that report values for the ordered moment $gS\sim 2\mu_{B}$ (reported in Ref. \onlinecite{Rodriguez11:84} to be 1.78(3) $\mu_{B}$/Fe for this same Fe$_{1.057(7)}$Te sample based on neutron powder diffraction).  With the Lande factor $g \equiv 2$ this implies a net $S\sim$ 1, implying that the magnetic Fe$^{2+}$ ion in Fe$_{1+x}$Te is in the strong crystal field limit with 4 electrons fully occupying the lower $|e_{g}\rangle$ and 2 in the triply degenerate and higher energy $|t_{g}\rangle$ manifold.  This scenario is further supported through consideration of total moment sum rules~\cite{Fobes14:112,Stock14:90} of neutron scattering that report significantly reduced magnetic spectral weight over what would be predicted from a $S=2$ (weak crystal field) ground state with a singlet orbital ground state.  The strong crystal field magnetic $S=1$ ground state has an underlying triplet orbital degeneracy (that can be described by an effective $l=1$).  Therefore magnetic spin-orbit coupling is a relevant term in the magnetic Hamiltonian with $\mathcal{H}_{SO}=\lambda \vec{l} \cdot \vec{S}$ in this scenario.

As is clear from the crystallography, the crystalline electric field environment surrounding the Fe$^{2+}$ ion is not perfectly tetrahedral and is distorted introducing a competing energy scale to spin-orbit coupling characterized by the distortion.  Considering the simplest case with a dominant axis for the distortion relevant in the spin-orbital Hamiltonian would have the approximate form $\mathcal{H}_{dis}=\Gamma (l_{z}^2-{2\over3})$ with $\Gamma$ quantifying the energetic size of this term.   This distortion ultimately breaks the orbital degeneracy at the structural transition.   The relative sizes of the distortion term to spin-orbit coupling (characterized by the ratio $\Gamma/\lambda$) are important for the nature of the structural and magnetic transitions at low temperatures. When the structural distortion energetics are dominant ($\Gamma/\lambda \gg 1$), we would anticipate a structural distortion at high temperatures with a lower temperature magnetic transition.  Alternatively, in the case that spin-orbit coupling is the dominant energy scale ($\Gamma/\lambda \ll 1$), given that the orbitals are coupled to the structure (via spin-orbit coupling) it would be expected that both magnetic and structural transitions would occur at the same temperature.  Such a competition between distortion and spin-orbit energetics has been modeled and compared to neutron spectroscopy data in insulating CoO~\cite{Sales09:79} and MgV$_{2}$O$_{4}$~\cite{Lane23:5}.

Given the importance of local magnetism in Fe$_{1+x}$Te and the work done on the FeSe counterpart illustrating the role that orbital fluctuations play in driving the structural and nematic orders, we consider our results in terms of this simplified localized picture.  Since the electronic behavior in Fe$_{1+x}$Te, FeSe, and the pnictides display many similarities, any interpretation of the critical properties of these materials must provide a consistent picture across the different classes of iron based materials.  Given that magnetic and structural distortions in Fe$_{1.057(7)}$Te occur at the same temperature (confirmed here with x-ray diffraction and spherical polarimetry), this would indicate that spin-orbit coupling is the dominant ($\Gamma/\lambda \ll 1$) energy scale with FeSe (and its nematic phase) representing the case of ($\Gamma/\lambda \gg 1$).  This would also be consistent with the structural nematic order being observed at the higher temperature of $\sim$ 90 K in FeSe, implying a larger energy scale for the distortion parameters over the fixed spin-orbit energy scales.  This might also be consistent with the shorter lattice constant in FeSe which would inevitably increase crystalline electric distortion terms in the magnetic Hamiltonian.  We, therefore, speculate that the relative strength of distortion to spin-orbit coupling energy terms in the magnetic Hamiltonian represents a control parameter for stabilizing nematic order in chalcogenides and pnictides.  

\section{Summary}

We have measured the temperature dependence of the soft $C_{66}$ elastic constant near the magneto-structural distortion in Fe$_{1.057(7)}$Te where it enters a low temperature monoclinic unit cell and a bicollinear magnetic structure.  This elastic constant has been found to be sensitive to nematic correlations in other iron based systems, however, the Fe$_{1+x}$Te series of compounds does not display nematic phases.  We speculate that the relative size of spin-orbital to distortion energetics is key for stabilizing nematic correlations and suggest Fe$_{1+x}$Te represents a case where spin-orbit terms are dominant in the magnetic Hamiltonian.

\begin{acknowledgements}
	
	We would like to thank C. Kirk and R. Rae for their help in using the Smartlab Rigaku facility and N. Giles-Donovan for helpful discussions. We are grateful for funding from the EPSRC and the STFC.
	
\end{acknowledgements}


%

\end{document}